# Effects of VR Gaming and Game Genre on Player Experience

Michael Carroll
*Department of Computer Science*
*State University of New York at Oswego*
Oswego, NY, USA
mcarroll@oswego.edu

Ethan Osborne
*Department of Computer Science*
*State University of New York at Oswego*
Oswego, NY, USA
eosborn2@oswego.edu

Caglar Yildirim[*]
*Department of Computer Science*
*State University of New York at Oswego*
Oswego, NY USA
caglar.yildirim@oswego.edu

*Abstract*— **With the increasing availability of modern virtual reality (VR) headsets, the use and applications of VR technology for gaming purposes have become more pervasive than ever. Despite the growing popularity of VR gaming, user studies into how it might affect the player experience (PX) during the gameplay are scarce. Accordingly, the current study investigated the effects of VR gaming and game genre on PX. We compared PX metrics for two game genres, strategy (more interactive) and racing (less interactive), across two gaming platforms, VR and traditional desktop gaming. Participants were randomly assigned to one of the gaming platforms, played both a strategy and racing game on their corresponding platform, and provided PX ratings. Results revealed that, regardless of the game genre, participants in the VR gaming condition experienced a greater level of sense of presence than did those in the desktop gaming condition. That said, results showed that the two gaming platforms did not significantly differ from one another in PX ratings. As for the effect of game genre, participants provided greater PX ratings for the strategy game than for the racing game, regardless of whether the game was played on a VR headset or desktop computer. Collectively, these results indicate that although VR gaming affords a greater sense of presence in the game environment, this increase in presence does not seem to translate into a more satisfactory PX when playing either a strategy or racing game.**

## I. Introduction

The way in which a player interacts with a game can alter their experience with the game. With the rise of virtual reality (VR) as a commercial product, the question developed of just how this technology can impact the gaming experience. The Oculus Rift was one of the first head mounted displays (HMDs) to become commercially available to users and paved the way for the VR boom that we are seeing today [1]. Even with the negative side effects associated with VR, mainly cybersickness, the uptake in the technology has been swift in recent years [1]. The biggest difference in VR and traditional desktop gaming is the use of an HMD through which the player sees the virtual environment. The use of an HMD enables the user to control the point of view (POV) with their head and freely look around their environment, which is not possible with desktop gaming [2]. This distinct difference is what many VR companies and developers say enhance the gaming experience of the platform. With many VR systems promising a more immersive and better gaming experience, the question arises as to whether that really is the case.

Some variables included in VR that increase player experience (PX) are immersive qualities and game genre. These factors, among others, are indicative of a game that should be rated with a high PX. As it currently stands most research into VR and gaming experience has focused on first-person shooter games [3, 4]. However, as more and more games are released for VR, it is important to consider how other game genres affect the PX during the VR gameplay.

### A. VR Gaming and Factors Affecting PX

Gaming in a virtual environment (VE) introduces new elements to the game play that designers and developers need to consider, including immersion and presence. These elements have helped to define what is considered the PX in VR and can be what propels a game to the status of Red Dead Redemption, a 2010 Western themed open-world RPG that was critically acclaimed or ends the game's "life".

Several theories have been posited to describe what contributes to a player's enjoyment of a video game. One such theory, the schema theory, states that there are factors and pleasures behind the enjoyment of the game [5]. Some of these factors include effectance, escape, and achieving a sense of flow, amongst others [5]. Effectance can be defined as a player feeling empowered by completing an action and having the game respond to that action. Escape is is considered to be

---

achieved when a player supplants themselves into an alternate reality fictional world present in a game [5]. A sense of flow is achieved when the player feels like they are completely in the game world. All of these factors, among others, facilitate the immersion of users into the gameworld and increase enjoyment.

Immersion in video games is the sense that the player is completely lost in the VE due to the quality of the sensory feedback provided by the VE. According to Mestre, immersion is achieved when sensations experienced in the VE overtake real world sensations [6]. This is important in VR gaming, as a break in immersion can cause the player to have a mediocre experience. For instance, Slater, Linackis, Usoh, and Kooper [7] looked at the difference in immersion while playing tri-dimensional chess between players wearing an HMD or playing on a TV screen. The participants were then placed in one of two environments: a garden environment or an empty environment, which was described by the experimenters as a void. Results showed that the HMD condition contributed to a better performance in both environments, and that being in a garden environment helped to conduce better PX than it did in the neutral environment. This indicates that the more salient the VE is, the higher the likelihood a better PX is. This relates to the aforementioned state of flow; if a player is placed in an environment that offers the ability to be in a different world (in a garden playing chess) they are more likely to give into that notion that they are in fact in a garden. This "notion" can be described as presence, which is yet another factor that contributes to PX during gameplay.

There has not been one accepted definition of presence because it is a hard measure to define. Rather, there have been many attempts by scholars to define it—usually within the constraints of the experiment they are running. Slater et al. [7] define presence as a construct that refers to the extent to which a person's behavior in the VE is representative of their behavior in comparable situations in real life, rather than with how well they perform as such. As the authors here argue, presence is different from immersion in that a player does not need to perform well in the VE to feel present. Rather, the VE has to "perform" well to allow the player to feel present.

Slater and Usoh [8] compared participants using an HMD in a VE to those who were navigating the same environment through an input device (mouse) and a TV. The environment was comprised of six doors in a hallway, with each door corresponding to a different scenario related to the VE-person interaction [8]. Results of this study indicate that there are exogenous factors that are capable of both causing a greater sense of presence and of taking the presence away. One of these exogenous factors contributing to the sense of presence was interacting with objects/doing a task. Interacting with objects increases the salience of the environment, which could allow the participants to achieve a flow state; previously stated to increase immersion and presence within the environment. It is believed that the more interactive a game is, the higher the likelihood the player will experience a greater sense of presence.

*B. Effects of Gaming Platform*

Previous research has actively investigated the putative effects of gaming platform on PX, mainly focusing on the comparison between VR and desktop gaming platforms. One study in particular by Shelstad, Smith, and Chaparro explored how VR affected the gaming experience of the strategy game *DG2: Defense Grid 2* [16]. Shelstad et al. found that the VR version of the game was rated as more enjoyable by the participants than that of the desktop version [1]. The idea behind VR providing a better gaming experience is that users are able to feel as though they are truly apart of their environment, a feeling that is hard to imitate in traditional desktop gaming. Another study into the viability of VR as a more enjoyable gaming platform showed that users found the use of VR in gaming more appealing than the likes of a 2D monitor, which in this study was a tablet [9]. On the contrary, other research has shown that there does not seem to be much a difference between VR and traditional desktop gaming, as was the case with the study by Yildirim, Carroll, Hufnal, Johnson, Pericles [2]. The key difference between these different studies was the genre of the game that was used in the study. Given the mixed findings regarding the effects of VR gaming on PX, it is important to further study this topic to better understand the influence of gaming platform on PX.

*C. Effects of Game Genre*

Extant literature on the effects of game genre and VR gaming has focused on how different games might impact immersion and sense of presence in the game environment. One study compared a fighting game, a shooter game, and a racing game over two different platforms, VR and PC with a joystick [3]. Predictions of this experiment were that presence would be greater overall in VR compared to PC and would be the greatest in the fighting game. The researchers were also interested in whether playing against another person would increase presence as opposed to playing against a computer. Results indicated that participants felt most present in the console as opposed to the VR condition and that the highest sense of presence came from the shooting game as opposed to the fighting game. There was also no significant effect of player against player compared to player against computer. The authors argued that participants' comfortability and familiarity with standard console play could have contributed to the higher levels of presence. This study indicates that further research is warranted to further elucidate the relationship between VR gaming and game genre and their effects on presence.

First person shooter (FPS) games have been utilized in previous research due to their perceived higher levels of immersion in a VR environment [2]. The higher levels of immersion were thought to bring about a higher PX as well. Results of this study indicated that there was no difference in presence levels or GUESS scores between VR and the traditional desktop experience. This suggests that FPS games might not be the best genre for immersion in VR, because FPS games are immersive enough in a traditional gaming experience [2]. Conversely, strategy games have been shown to elicit a greater PX in VR as opposed to the traditional experience [1].

The current study was an attempt to replicate Shelstad et al. [1] as well as continue the research conducted by Yildirim et al. [2]. The purpose of the current study was to examine the effects of game platform and game genre on PX and presence. The following are the hypotheses of the current study:

H1a: Participants would experience a greater sense of presence in the VR gaming platform, when compared to the desktop gaming platform.

H1b: Participants would experience a greater sense of presence in the strategy game, when compared to the racing game.

H1c: The effect of VR gaming on sense of presence would be greater in the strategy game, compared to the racing game.

H2a: Participants would experience a greater PX in the VR gaming platform, when compared to the desktop gaming platform.

H2b: Participants would experience a greater PX in the strategy game, when compared to the racing game.

H2c: The effect of VR gaming on PX would be greater in the strategy game compared to the racing game.

## II. METHOD

### A. Design

This study used a two-factorial (gaming platform x game genre) mixed design. The independent variables were the type of gaming platform, Oculus Rift and Desktop PC, which was manipulated between-subjects, and the genre of the game that was played, Strategy and Racing, which was manipulated within-subjects. The dependent variables were PX, as measured by the Game User Experience Satisfaction Scale (GUESS) and sense of presence, as measured with the Presence Scale.

### B. Participants

The participants of this study were recruited from the college student population at the university where this study was conducted. In total there were 36 participants, with 18 in each condition. The sample consisted of 30 female and 16 male participants. The average age of the participants was 22.31 (*SD* = 3.98). Participants were recruited via email announcements and were offered the opportunity to enter a raffle to win one of the three $10 gift cards.

### C. Materials

*1) Game User Satisfaction Scale*

The Game User Satisfaction Scale (GUESS) is a multidimensional measure of the user's PX. The GUESS measures 9 different aspects of the gaming experience including: Usability/Playability, Narratives, Play Engrossment, Enjoyment, Creative Freedom, Audio Aesthetics, Personal Gratification, Social Connectivity, and Visual Aesthetics [10]. With the nature of the games that were used in this study being focused on racing and strategy both the narrative and social connectivity aspects of this measure were omitted from the survey that was given to the participants. Overall the GUESS scale encompasses 55 items used in the measurement, but for the purposes of this study only 44 were deployed [10]. This study used the remaining seven dimensions averaged together to gather the overall GUESS score. With the GUESS measure, a higher score indicated a more satisfactory PX.

*2) Presence Scale*

In order to determine the level of presence the participants felt, the presence scale was used [11]. The presence scale contains eight items, all of which are on a 7-point Likert scale. These measures are all self-reported and the results are averaged to determine the level of presence the individual experiences. The higher the score on the presence scale, the more present the individual feels during their experience.

*3) Gaming Platforms*

This study used two different gaming platforms, one VR headset and a Desktop (Windows) PC (see Fig. 1 and Fig. 2). The VR headset that was used in this study was the Oculus Rift [12]. The Oculus Rift is a VR headset that makes use of a HMD to track the users position around the VE. The Oculus Rift features a display with a 1080 x 1200 resolution per eye with a refresh rate of 90 Hz [13]. This HMD also features a gyroscope and accelerometer, which is used to track the user's head motion. The Oculus Rift also has a field of view (FOV) of 110 degrees [14]. With the use of the Oculus Rift comes the ability to use motion controllers, but for the purposes of this study participants were given an Xbox controller to use. This was done to keep the input type a constant between the platforms, as the desktop versions of the games are not compatible with the motion controls.

*4) Game Genres*

This study explored the effects of two different game genres: Racing and Strategy. The racing game what was used during this study was *Assetto Corsa* [15]. *Assetto Corsa* is a racing simulator game where the player uses a first-person perspective to navigate a car around a track. Both of the VR and desktop versions of the game were identical.

The second game that was played during this study was the strategy game *DG2: Defense Grid 2* [16]. *DG2: Defense Grid 2* is a top-down tower defense game, where the player aims to stop enemies from depleting their cores through the use of towers that attack the enemies as they maneuver through the map. Similar to *Assetto Corsa,* there is no differences in the versions of the game between VR and desktop platforms. Participants within the study played both games during the course of the experiment in a counterbalanced order.

### D. Procedure

When participants first arrived, they were greeted and asked to read over the informed consent form. While the participants were reviewing the form the experimenter randomly drew the participants starting condition, which were PC Racing, PC Strategy, VR Racing, or VR Strategy. This selection determined the platform on which the participant played both game genres as well as which game would be played first. Once the starting condition was selected participants were given a short demo (using the Xbox controller) of the first game.

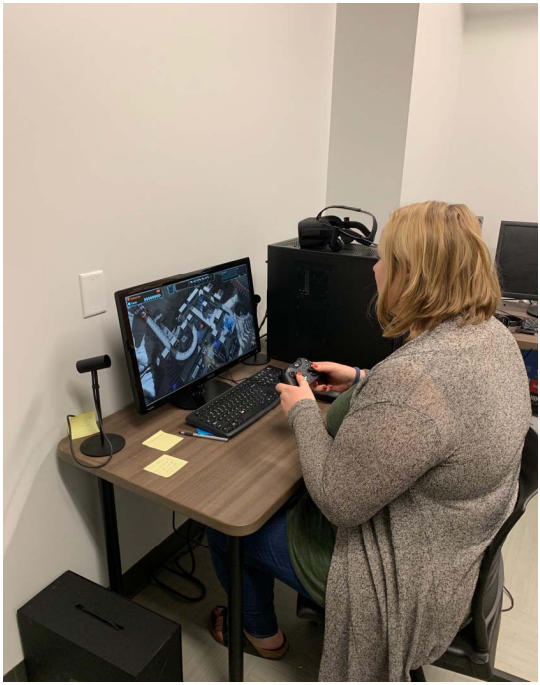

**Fig. 1.** Image of participant in Desktop condition.

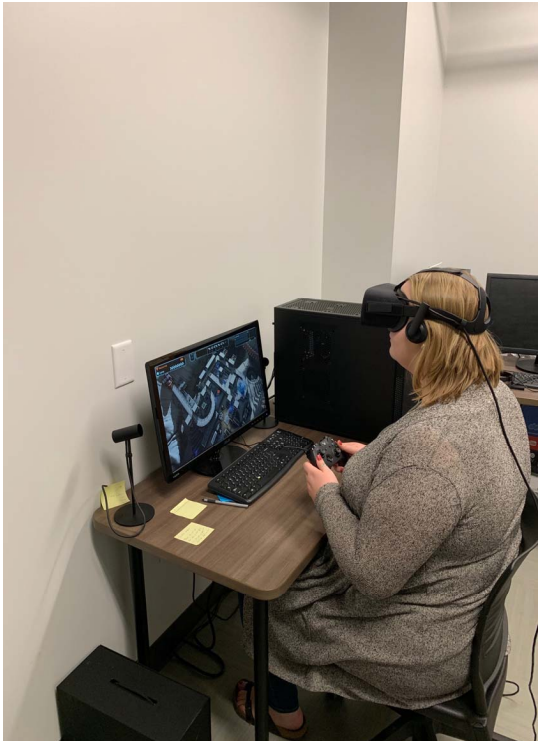

**Fig. 2**. Image of participant in VR condition.

Following the random assignment of the starting condition, the experimenter left the room and the participants were left to play the game for 10 minutes. After 10 minutes had passed the experimenter re-entered the room and directed the participants to the surveys. The experimenter once again left the room whilst the participants completed the surveys. Upon conclusion of the first set of surveys participants were given a short break between the games, while the experimenter prepared the next part of the experiment. After the break participants were brought back into the room and given a short demo (using the Xbox controller) on the next game they would play. Once again, they were given 10 minutes to play the game while the experimenter waited outside. After 10 minutes had passed the experimenter re-entered the room and guided the participant to the final set of surveys. Once the participant had concluded the final surveys they were debriefed and afforded the opportunity to ask any questions. The entire experiment was completed in less than 60 minutes.

## III. RESULTS

Several descriptive and inferential statistics tests were run to test the hypotheses of the current study. Table 1 and Table 2 present the descriptive statistics and the results of inferential statistics tests for each of the dependent variables, respectively.

Table 1. Descriptive Statistics for the Dependent Variables

|  | **Strategy** | **Racing** |
|---|---|---|
|  | *M* (*SD*) | *M* (*SD*) |
| **Sense of Presence** |  |  |
| Desktop PC | 3.63 (1.43) | 2.88 (1.62) |
| VR | 4.65 (1.05) | 3.94 (.984) |
| **Player Experience** |  |  |
| Desktop PC | 35.10 (7.00) | 29.20 (9.16) |
| VR | 37.06 (6.25) | 29.32 (7.40) |

Sense of Presence was measured by the average score on PQ. Player Experience denotes PX ratings as indexed by the total score on GUESS.

Table 2. Results of Factorial ANOVA

| Variable | $F(1, 34)$ | $p$ | $\eta^2$ |
|---|---|---|---|
| **Sense of Presence** |  |  |  |
| Platform | 8.10 | .007 | .192 |
| Game Genre | 10.05 | .003 | .228 |
| Platform x Genre | .015 | .905 | .0 |
| **Player Experience** |  |  |  |
| Platform | .229 | .636 | .964 |
| Game Genre | 29.18 | <.001 | .462 |
| Platform x Genre | .534 | .470 | .015 |

$\eta^2$ represents partial eta-squared values as an estimate of effect size. Alpha level set at .05 for all hypothesis tests.

A 2 (gaming platform) x 2 (game genre) mixed ANOVA was conducted to examine the effect of gaming platform (VR vs. desktop gaming) and game genre (strategy vs. racing game) on sense of presence levels (Fig. 3). In contrast with Hypothesis 1c, there was no significant interaction effect between gaming platform and game genre, $F(1, 34) = .015$, $p = .905$, indicating the effect of gaming platform on sense of presence did not change as a function of the game played. Therefore, each main effect was individually examined.

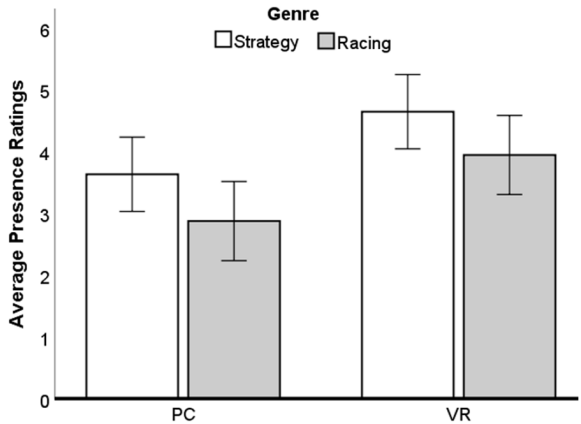

**Fig. 3**: Sense of presence levels

Results revealed a significant main effect of gaming platform on sense of presence, $F(1, 34) = 8.10, p = .007$ (Hypothesis 1a). This indicates that regardless of which game was played, participants experienced a greater sense of presence when playing the game on the VR headset than on the desktop computer. Similarly, there was a main effect of game genre on sense of presence, $F(1, 34) = 10.05, p = .003$ (Hypothesis 1c). This indicates that regardless of whether the game was played in VR, participants experienced a greater sense of presence when playing the strategy game, compared to the racing game.

As for the effect of gaming platform and game genre on PX (Fig. 4), results of ANOVA revealed that there was no significant interaction between these two factors, $F(1, 34) = .534, p = .47$ (Hypothesis 2c), indicating the effect of gaming platform on sense of presence did not change as a function of the game played. Therefore, each main effect was individually examined. Results showed no significant main effect of gaming platform on PX ratings, $F(1, 34) = .229, p = .636$ (Hypothesis 2a), suggesting that the game platform used to play the game did not significantly influence PX, regardless of the game played. That said, the main effect of game genre on PX ratings was statistically significant, $F(1, 34) = 29.18, p < .001$ (Hypothesis 2b). This indicates that regardless of whether the game was played in VR, participants experienced a greater PX when playing the strategy game, compared to the racing game.

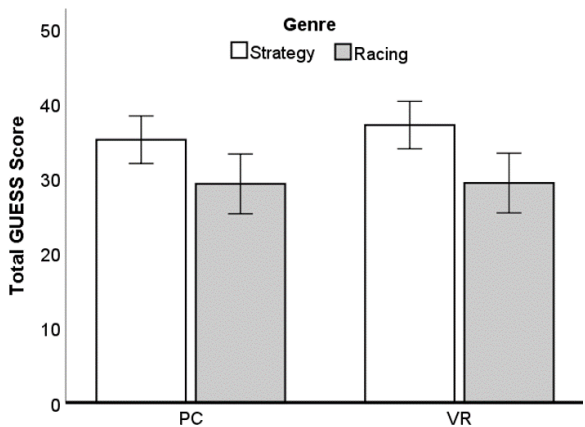

**Fig. 4**. PX ratings as a function of gaming platform and genre

## IV. DISCUSSION

The purpose of this study was to investigate VR and traditional desktop (PC) gaming to determine which resulted in a greater PX, and to determine if game genre had an effect on PX as well. The hypothesis of this study was that VR gaming would provide a greater PX than traditional desktop (PC) gaming. Results of this experiment indicated that there was no significant differences between VR gaming and desktop (PC) gaming in terms of PX. This differs from previous research demonstrating that VR gaming afforded a greater PX compared to desktop (PC) gaming [1]. Since this study was not a one-to-one replication of Shelstad et al. [1], it can be inferred that the differences in findings resulted from the differences in the studies. The previous work focused solely on the strategy game whereas this study included a racing game as well to check for differences in game genre. Also, the previous work was a within-subjects design as opposed to the current studies between-subjects design, which could have contributed to a greater difference in the scores.

Even though there were no significant results between game platforms, there was a significant difference of PX between game genres. Results indicated that the strategy game brought a higher PX than did the racing game. This can be attributed to the fact that the strategy game used in the current study was more interactive than was the racing game. During the racing game, participants drove around a track with no other cars for the ten-minute period. During the strategy game, participants were constantly being stimulated by the game due to the constant need to defend the towers. These results are consistent with those found by Lindley and Sennersten [5], in which participants achieved a higher sense of effectance during the strategy game due to the constant feedback loops granted to them for completing actions. Therefore, it can be inferred that regardless of platform, more interactivity can lead to a greater PX.

One aspect of VR and the PX that was explored in this study was how both the platform and game genre affected the sense of presence. As was predicted in this study, the level of presence was greater in the VR condition compared to the desktop (PC) condition, regardless of the genre of the game played. There have been many studies that have focused on how interaction with VEs can affect the user's feeling of presence [12, 19]. In their study of presence Witmer and Singer [11] highlighted the importance of interactivity in VR on the sense of presence in the VE. Witmer and Singer [11] also discuss how the better the interaction with a VE is the better sense of presence the user will feel. Along similar lines as Witmer and Singer [11] is the research of Schubert, Friedmann, and Regenbrecht [17], who's research determined that the level of interaction with a VE is one of the main predictors, the other of which was immersion, of presence level. The results of the current study are consistent with previous research in that the game with more interactions produced a greater level of sense of presence during the gameplay.

While the results of presence levels in the current study are consistent with previous research, it does differ from the results of the study by Yildirim et al. [2] This research did not find a significant difference in presence [2]. Since the current study is a follow up to Yildirim et al.'s [2] work it is important to note that the current study was able to produce a difference in presence levels, which was not achieved previously. It is possible that changes in interaction style and using different game genres, which also have different interactions, contributed to produce this difference. Yildirim et al. [2] used a first-person shooter game and compared player experience and presence across two VR conditions and desktop (PC) gaming. The authors found no differences across these three conditions in PX and presence levels. Therefore, it could be argued that VR gaming and traditional desktop gaming do not differ in PX and sense of presence levels when the game is a first-person shooter game, a game genre involving fast paced, highly interactive combat environments. That said, based on the findings of the current study, it is seen that VR gaming affords a greater sense of presence when playing a strategy and racing game. Since the current study did not include a first-person shooter game in the game genre comparison, future research is warranted to elucidate the extent to which this interpretation in tenable.

One of the limitations of the current study was that it relied on subjective measures to measure PX. There were no physiological measures taken due to technological issues while setting up the experiment, which prevented the use of these measures. Inclusion of physiological measures in the measurement of enjoyment while playing video games has the capacity to further elucidate the effects that game genre and platform can have on PX. Research has shown that physiological measures can be indicative of a greater PX [18]. This shows the efficacy of using physiological measures in research in this area and including physiological measures can help identify which game genre and platform is most engaging.

Overall, this study provided results showing that VR does not always afford a greater PX than does the traditional gaming experience. However, there are differences in game genres as they relate to PX; this could be due to the extent to which they fit into the schema theory [5]. As VR continues to become more popular commercially, it is important that research be done on it in order to curate the greatest PX. Future research into this topic should investigate a one-to-one replication of the Shelstad et al. study in order to replicate those results. Furthermore, other game genres must be tested to learn how effective they are at granting a good PX. The current study examined the roles of gaming platform and game genre on PX. Other factors, such as controller naturalness should be investigated as well to better understand what provides for an exquisite PX.

ACKNOWLEDGMENT

The study was conducted in the SUNY Oswego VR First Lab, established through the academic partnership between VR First and SUNY Oswego. The authors would like to thank VR First for supporting the establishment and maintenance of the lab.